\def\final{1}

\ifdefined\final
\documentclass[twocolumn,9pt]{article} 
\fi

\ifdefined\draft
\documentclass[onecolumn, 12pt]{elsarticle} 
\fi

\usepackage[square,numbers,sort&compress,comma]{natbib}

\usepackage{amsmath}
\usepackage{amssymb}
\usepackage{caption}
\usepackage{graphicx}
\usepackage{latexsym}
\usepackage{times}
\usepackage[pagewise]{lineno}
\usepackage{hyperref}

\topmargin - 12pt 
\ifdefined\final

\makeatletter
\renewcommand{\maketitle}{%
  \begin{center}%
    {\LARGE \@title \par}%
    \vskip 1em%
    {\large \lineskip .5em \@author \par}%
  \end{center}%
  \par \vskip 0.5em}
\makeatother

\renewenvironment{abstract}%
              {
               \small
               {\bfseries \abstractname}
               \par
               \vspace{10pt}
              }
              {}

\renewcommand\abstractname{Abstract}

\newcommand{\nomenclature}
              [1]
              {
               \bgroup
               \flushleft
               \small\bf
               #1
               \par
               \egroup
              }

\renewcommand{\section}
              [1]
              {
               \bgroup
               \flushleft
               \small\bf
               \refstepcounter{section}
               \arabic{section}. #1
               \par
               \egroup
              }

\renewcommand{\subsection}
              [1]
              {
               \bgroup
               \flushleft
               \small\em
               \refstepcounter{subsection}
               \arabic{section}.
               \arabic{subsection}. #1
               \par
               \egroup
              }

\renewcommand{\subsubsection}
              [1]
              {
               \bgroup
               \flushleft
               \small\em
               \refstepcounter{subsubsection}
               \arabic{section}.
               \arabic{subsection}.
               \arabic{subsubsection}. #1
               \par
               \egroup
              }
\fi

  \newcommand{\acknowledgement}
              [1]
              {
               \bgroup
               \flushleft
               \small\bf
               #1
               \par
               \egroup
              }

\ifdefined\final
  \newcommand{\sectionbib}
              [1]
              {
               \bgroup
               \flushleft
               \small\bf
               #1
               \par
               \egroup
              }

\fi

\ifdefined\draft

\fi

\graphicspath{{./figures/}}

\begin{document}



\small
\ifdefined\final
\baselineskip 10pt
\fi

\setcounter{page}{1}
\title{\LARGE \bf High-resolution \emph{in situ} analysis of biomass pyrolysis by combining quantitative synchrotron $\mu$CT and 3D particle-resolved simulations}

\author{{\large Emeric~Boigné$^{a,*}$, Mohamed~M.~Ahmed$^{b}$, Collin~Foster$^{c,d}$, Edna~R.~Toro$^{a}$, \\
Dilworth~Y.~Parkinson$^{d}$, Alastair~A.~MacDowell$^{d}$, Harold~S.~Barnard$^{d}$, \\
Francesco~Panerai$^{c}$, Chiara~Saggese$^{b}$, Matthias~Ihme$^{a,e,f}$}\\[10pt]
    {\footnotesize \em $^a$Department of Mechanical Engineering, Stanford University, Stanford CA 94305, USA.}\\[-5pt]
    {\footnotesize \em $^b$Materials Science Division, Lawrence Livermore National Laboratory, Livermore CA 94550, USA.}\\[-5pt]
    {\footnotesize \em $^c$Department of Aerospace Engineering, University of Illinois at Urbana-Champaign, Urbana IL 61801, USA.}\\[-5pt]
    {\footnotesize \em $^d$Advanced Light Source, Lawrence Berkeley National Laboratory, Berkeley CA 94720, USA.}\\[-5pt]
    {\footnotesize \em $^e$Department of Photon Science, SLAC National Accelerator Laboratory, Menlo Park CA 94025, USA.}\\[-5pt]
    {\footnotesize \em $^f$Department of Energy Science and Engineering, Stanford University, Stanford CA 94305, USA.}}

\date{}  

\ifdefined\final
\twocolumn[
\fi

\maketitle
\rule{\textwidth}{0.5pt}
\vspace{-5pt}

\ifdefined\final
\begin{abstract} 
\fi 

\ifdefined\draft
{\bf Abstract}
\fi

Understanding the coupling between transport, chemical kinetics, and structural response within solid fuel combustion requires \textit{in situ} measurements at high spatial resolution. To examine the dynamics of biomass pyrolysis, we optimize synchrotron X-ray micro-computed tomography ($\mu$CT), and conduct 3D particle-resolved simulations. A micro-focused heating cell with controlled heating and flow rate is employed, achieving temperatures of up to $1240$ K and 3D imaging at $3.24$ µm spatial and sub-minute temporal resolutions. The pyrolysis of three biomass materials at sample heating rates of $10$--$14$~K/min are examined. The $\mu$CT measurements capture pore deformation, cracking, anisotropic shrinkage, and provide simultaneous thermogravimetric and thermovolumetric analyses to examine specific secondary pyrolysis pathways. Complementary 3D simulations considering detailed kinetics and structural response reproduce the key trends observed, and identify deficiencies in capturing simultaneous mass and volume losses during pyrolysis. 

\ifdefined\final
\end{abstract}
\fi

\vspace{10pt}

{\bf Novelty and significance statement}

\vspace{10pt}
Despite increasingly detailed kinetic models~\cite{debiagi2018predictive, debiagi2024cellulose}, the pyrolysis of solid fuel particles remains poorly characterized because major uncertainties persist due to the difficulty of measuring reactions and structural deformations inside particles. Prior X-ray studies have probed inside fuel particles, but were restricted by low spatial resolution~\cite{boigne2020simultaneous, boigne2021structural}, poorly controlled heating rates~\cite{Ruhlandt2017}, or reduced contrast~\cite{Murai2020}, limiting analysis to qualitative observations. This work overcomes these limitations by optimizing $\mu$CT synchrotron experimental techniques, providing 3D \textit{in situ} measurements and simulations that directly resolve the pore-scale processes during pyrolysis. The results help understand pyromechanics~\cite{Lahouze2025}, intra-particle regime transitions~\cite{BrennanPecha2019, Pecha2021}, and provide quantitative evidence that supports specific secondary pyrolysis pathways. This work thus addresses key questions in solid biomass conversion~\cite{mettler2012top}: how to achieve accurate predictions, what drives shrinkage and matrix cracking, and how does pore-scale transport influence the exothermic secondary charring and tar-cracking reactions~\cite{debiagi2018predictive, Wang2022}.

\vspace{5pt}
\parbox{1.0\textwidth}{\footnotesize {\em Keywords:} \textit{In situ} X-ray $\mu$CT; Heterogeneous Combustion; Pyrolysis; Biomass; Particle-Resolved Simulations}
\rule{\textwidth}{0.5pt}
*Corresponding author.
\vspace{5pt}

\ifdefined\final
]
\fi

\section{Introduction\label{sec:introduction}} \addvspace{9pt}

Solid fuel chemistry is central to many processes, ranging from energy conversion and catalysis to biomass utilization and wildfire spotting~\cite{manzello2020role}. The multiscale nature of materials such as biomass, together with the multiphase nature of solid-fuel reactions, makes the coupling between anisotropic transport, heterogeneous kinetics, and structural response inherently complex. As a result, fundamental questions remain~\cite{mettler2012top}: How can heat transfer be accurately predicted in reacting porous particles? What mechanisms drive particle shrinkage, cracking, and anisotropic deformation? How does volatile transport influence overall heat release, given the strongly exothermic contribution of secondary charring and tar-cracking within pores~\cite{debiagi2018predictive, Wang2022}? 

Significant advances in reaction kinetics have improved our understanding of biomass devolatilization and secondary chemistry~\cite{Anca-Couce2016, debiagi2018predictive}. Yet, kinetics alone cannot describe the pore-scale transport and pyromechanics of reacting solids, which today remains largely inferred rather than measured.

X-ray micro-computed tomography ($\mu$CT) provides a means to overcome optical-access limitations and to retrieve realistic 3D fuel structures for simulations of biomass~\cite{Crowley2022} and coal~\cite{Fong2018, Liang2022}. In our previous work, we demonstrated the capabilities of CT laboratory sources to enable simultaneous measurements of gas temperature and solid density during biomass smoldering~\cite{boigne2020simultaneous, boigne2021structural}. However, these laboratory sources remain limited in spatio-temporal resolution, to $0.3$~mm and $90$~s in our setup. This restricts the analysis of rapid primary devolatilization, which occurs within $2$--$5$~min depending on heating rates.

To overcome limitations in photon flux, and spatio-temporal resolutions, recent studies have employed synchrotron $\mu$CT \emph{in situ} to examine transport-chemistry coupling regimes in carbon fiber oxidation~\cite{ringel2025carbon} and solid propellant~\cite{jiang2025heating}. Previous studies have also performed \emph{in situ} synchrotron $\mu$CT of biomass pyrolysis at $\mu$m resolution and scan times below $10$~s~\cite{Murai2020, Ruhlandt2017}. Yet, these biomass studies remain qualitative due to low contrast, limited heating rate control, or the need for phase contrast imaging~\cite{Ruhlandt2017}, thus prohibiting quantitative model comparisons.

To address these issues, the objective of this study is to enable quantitative measurements of pyrolysis by optimizing \emph{in situ} $\mu$CT, and conduct comparisons with detailed simulations. The key developments from this study include (i) a micro-focused heating cell enabling well-controlled operation in inert flow conditions at surface temperatures of up to $1240$~K, (ii) high-accuracy synchrotron $\mu$CT to retrieve quantitative time-resolved 3D measurements of solid density, (iii) quantitative processing of the $\mu$CT data to enable simultaneous thermogravimetric (TGA) and thermovolumetric (TVA) analyses, and detailed characterization of individual micro-pores. Specifically, to enable quantitative $\mu$CT measurements despite the low X-ray contrast of organic materials, the synchrotron $\mu$CT parameters are optimized by leveraging techniques originally developed for 3D pore-scale temperature measurements~\cite{boigne2022towards, Boigne2024}: minimization of white beam filtering, acceleration of $\mu$CT data acquisition rates, and optimization of contrast and temporal resolution trade-offs. These measurements are complemented with 3D simulations considering detailed chemical kinetics and structural responses.

\section{Methods\label{sec:methods}} \addvspace{10pt}

\subsection{Experimental method\label{sec:methods_exp}} \addvspace{10pt}

Experiments are conducted at the 8.3.2 $\mu$CT beamline of the Advanced Light Source synchrotron, operated in top-off mode with a constant ring current of $500$~mA. The X-ray beam is generated by a superconducting bending magnet of $4.37$~Tesla~\cite{Robin2005, mobilio2016synchrotron}. To maximize photon count and enable quantitative $\mu$CT measurements, the beamline is operated in white-light mode without monochromator, enhancing the photon count at the sample by two orders of magnitude. The beam is filtered only by four permanent beryllium vacuum windows (totaling $\approx 0.5$~mm Be) and a double-sided polished Si wafer of $4$~mm thickness~\cite{Boigne2024}. Minimal beam hardening is observed since biomass consists mainly of low-attenuation H, C, and O atoms. The sample-to-scintillator distance is set to $155$~mm, as a trade-off between minimizing heat load on the optics and reducing resolution from phase contrast effects. The pixel resolution is $3.24$~$\mu$m, and $1313$ images of $2560\times1400$ pixels are acquired for each scan using a pco.edge 5.5 CMOS camera~\cite{boigne2022towards} at $50$~frames per second, balancing azimuthal accuracy against dynamics resolution.

\begin{figure}[ht!]
\centering
\includegraphics[width=192pt]{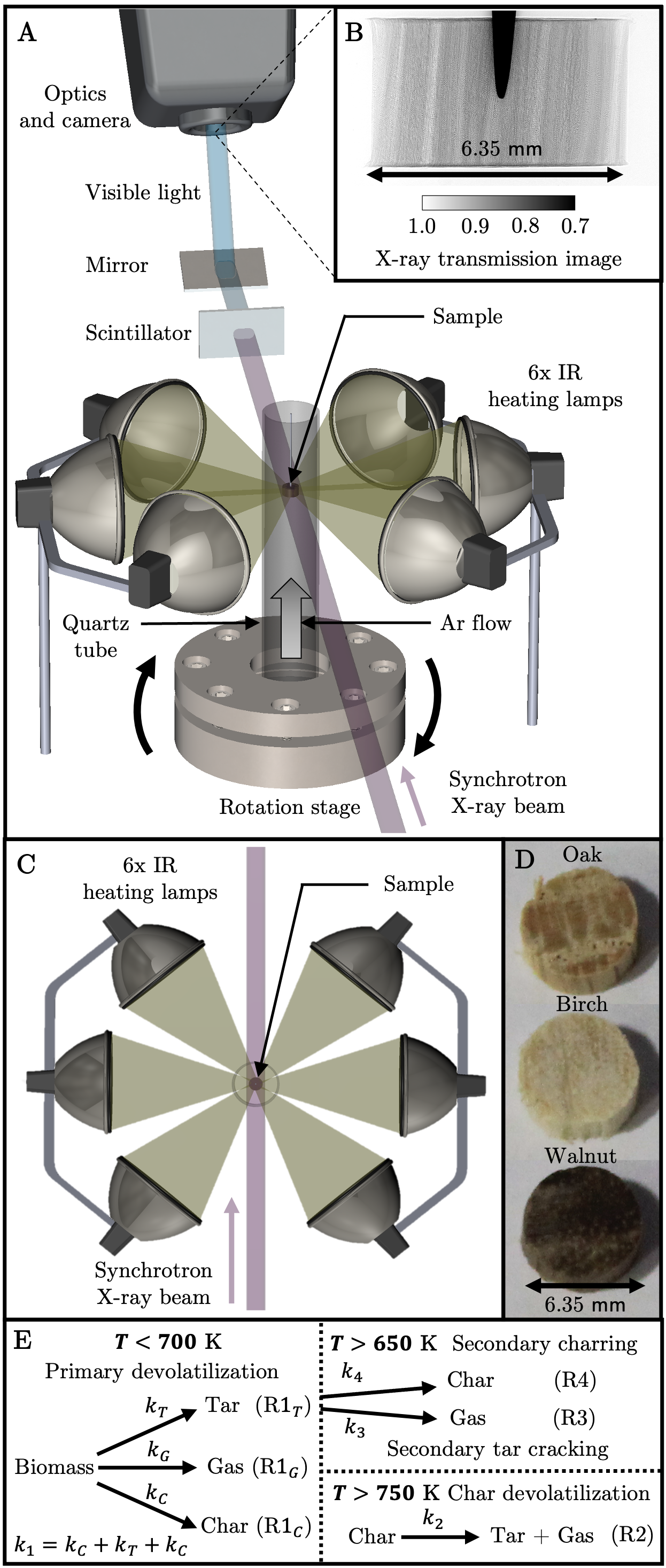}
\addvspace{2mm}\caption{\footnotesize Experimental setup to examine biomass pyrolysis. (\textbf{A}) Schematic of the experimental setup consisting of heating cell, synchrotron beam, and rotation stage. (\textbf{B}) One X-ray transmission image of pristine oak mounted on a steel needle ($3.24$~$\mu$m pixel size). (\textbf{C}) Top view of the heating system focused at the sample. (\textbf{D}) Pictures of three biomass samples. (\textbf{E}) Global chemical pathways of biomass pyrolysis, with representative temperature thresholds provided for the present discussion. But note that the underlying kinetics are complex and material-dependent~\cite{debiagi2018predictive, debiagi2024cellulose}.}
\label{FIG_SETUP}
\vspace{-0.1 in}
\end{figure}

Disc-shaped biomass samples, representative of firebrands~\cite{manzello2020role}, are examined using the micro-focused heating cell shown in Fig.~\ref{FIG_SETUP}. The heating cell is placed on the rotation stage of the synchrotron X-ray hutch. It contains six 150 W halogen lamps in a confocal arrangement, each with an ellipsoidal reflector focusing infrared radiation onto a 5 mm spot~\cite{Bale2012, Haboub2014, foster2023real}. Samples are held with a needle in a quartz tube of $20$~mm inner diameter, with an Ar flow of $4\pm0.1$ SLPM at atmospheric pressure metered by a mass flow controller from Alicat Scientific. We use samples ($6.35$ mm diameter, $3.3\pm0.2$ mm height, oven-dried at $383$ K for three hours) of red oak, walnut, and birch, respectively, representing ring-, semi-ring-, and diffuse-porous hardwoods. Proximate and ultimate analysis of the materials performed after drying are reported in a previous study~\cite{boigne2021structural}. The masses of the pristine and charred samples are measured within $0.2$~mg accuracy using a precision scale.

The heating lamps are held at $90$~W ($470$~K) for $300$~s before each run, after which heating begins ($t=0$). The power is increased linearly to $480$~W, at rates of $10$ and $14$~K/min with peak temperatures of $910$~K and $1240$~K, respectively. These rates were selected to operate in the complex regime of coupled transport-kinetic, while remaining compatible with the temporal resolution of the present $\mu$CT setup. Every $150$~s, X-ray images are acquired while continuously rotating the sample 360° over $52.5$~s, producing two 180° scans. The lamps and exhaust remain fixed while the quartz tube with the sample rotates. Bright field images are acquired before every revolution by moving the sample upwards such that the field-of-view only contains a section of the quartz tube without sample. After heating, lamps are turned off and scans continue for 20~minutes during cooling.

\begin{figure}[h!t]
\centering
\includegraphics[width=192pt]{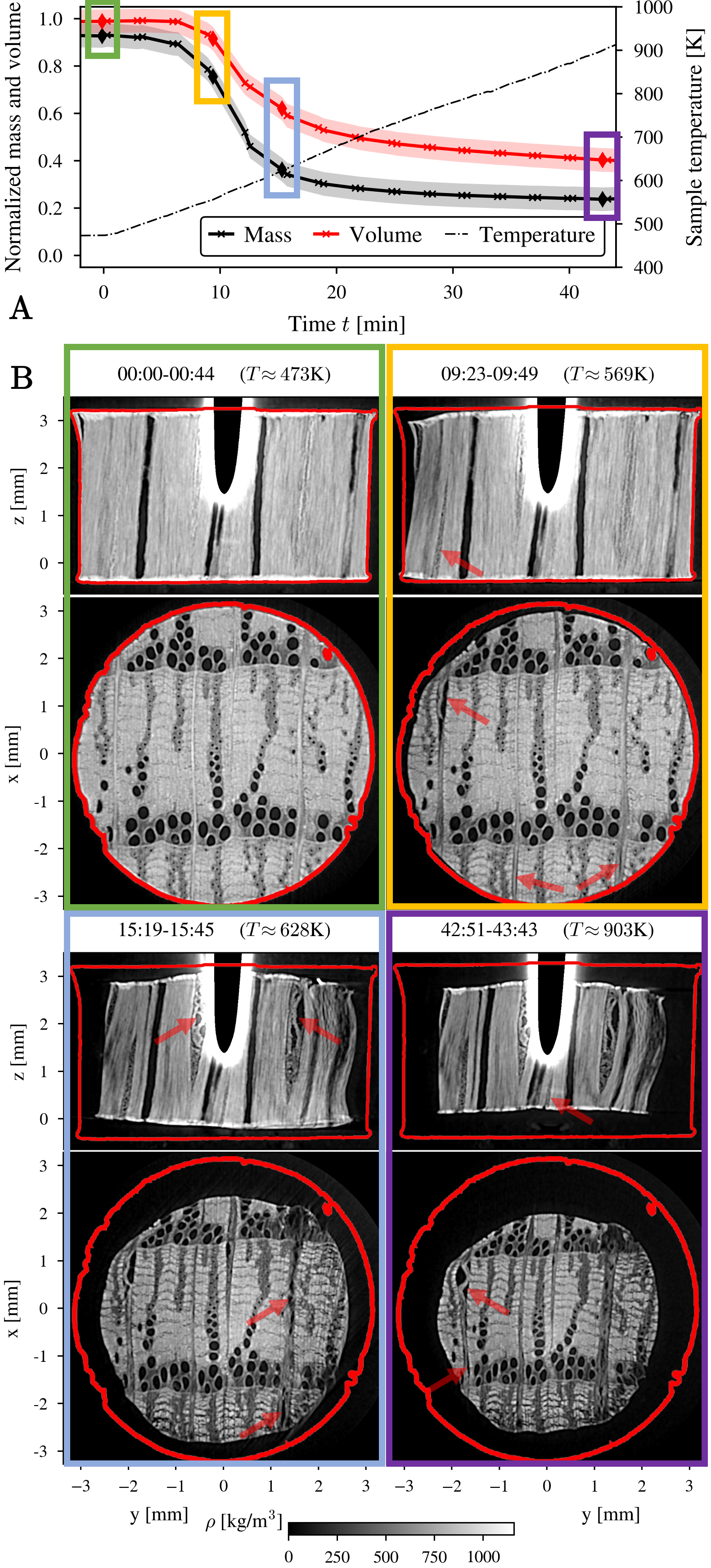}
\addvspace{2mm}\caption{\footnotesize $\mu$CT analysis of oak sample mounted on needle and heated at $10$~K/min. (\textbf{A}) Temperature readings from thermocouple calibration during pyrolysis, and measurements of sample mass and volume evaluated via $\mu$CT with shadings indicating experimental uncertainties. (\textbf{B}) Cross-sections of the $\mu$CT measurements of density at different times color-coded. The initial sample surface is delineated with red contours, and red arrows point to crack formations.}
\label{FIG_OAK}
\vspace{-0.1 in}
\end{figure}

The temperature is calibrated in dedicated runs with a K-type thermocouple within a sheath of $1$~mm diameter Inconel~$600$ positioned on the upper surface of oak samples. It is rated for temperature readings of up to $1400$~K by manufacturer TC Direct, Hillside, IL, USA (thermocouple reference 208-110). To avoid image artifacts from scattering of the X-rays on the metallic probe, the thermocouple is used only during calibration. The thermocouple measurements show that the heating rates are slow enough to obtain a nearly linear response in sample temperature  (Fig.~\ref{FIG_OAK}\textbf{A}). All temperature values reported in the study are from these calibration experiments, including the $10$ and $14$~K/min heating rates and the corresponding maximum temperatures of $910$~K and $1240$~K. 

Each 180° dataset is reconstructed by filtered back-projection (ASTRA Toolbox~\cite{VanAarle2016}), to retrieve about sixty full 3D volumes per sample. The $\mu$CT center-of-rotation is manually adjusted within $0.1$~pixel for each dataset. The method used to quantitatively measure solid density from the raw $\mu$CT signal is detailed in previous work~\cite{boigne2021structural}. 3D datasets are binned by a factor of two for a voxel size of $6.48$~$\mu$m, and individually aligned with SimpleITK~\cite{yaniv2018simpleitk}. The aligned datasets are then filtered using a 3D Gaussian smoothing kernel with standard deviation of $\sigma=3$ pixels to segment between the solid and porous phase via a global time-varying threshold. Pores larger than 10--20~$\mu$m are individually characterized using the particle analyzer and skeletonization functions in scikit-image. The GPU-optimized scripts for the $\mu$CT reconstruction with the ASTRA Toolbox, data clean-up with PyTorch, and alignment with SimpleITK are available online (\url{https://github.com/IhmeGroup/tomoTools_als832}).

Two sample volumes are calculated after segmentation: the micro-dense sample volume $V_{\mu \text{m}}$ excludes the resolved $10$--$20$~$\mu$m-diameter pores, while the bulk sample volume $V$ contains the volume occupied by the micro-pores. The porosity resolved at the micro-scale is then computed as $\varepsilon_{\mu \text{m}} = 1-V_{\mu \text{m}}/V$. The sample height $H$ along the vertical grain direction is calculated over time by measuring the minimal distance separating two horizontal planes placed above and below the sample. The equivalent sample diameter $D$ is evaluated by assuming a cylindrical sample shape with $V=\pi H D^2/4$. The skeleton of each pore vessel is extracted to compute the pore tortuosity, which is evaluated as the ratio $\tau= L_s/h_p$ of the linear skeleton length $L_s$ to the pore height along the grain direction $h_p$. The pore eccentricity and equivalent diameter are obtained as averages over the vertical height of each pore. Specifically, at each horizontal slice, an ellipsoid is fitted to the 2D contour of each pore, from which local values of eccentricity and equivalent diameter are computed. These slice-specific values are then averaged along the pore height to
obtain mean values for each pore.

\begin{table*}[ht!] 
\begin{center}
\begin{footnotesize}
\centering
\caption{Biomass physical properties used in the simulations and characteristic heating time-scales. The solid densities are the dense-material values which are obtained from measurements of bulk density and porosity (see supplementary Table~S1). Values from the literature are used for the darcy tensor~\cite{debiagi2018predictive}, the $\lambda^s$ thermal conductivity~\cite{forest1987wood, flity2024thermal}, and the biomass and char heat capacities~\cite{Gentile2017}. The Biot number is evaluated as $\textrm{Bi} = hL/\lambda^s$, with $h=80$~W/(m$^2$K) the effective heat transfer coefficient evaluated from the thermocouple calibrations (see supplementary Fig.~S1), and $H=3.3$~mm the sample length-scale.}
\label{TAB_THERMO_PROP}
\centering
\begin{tabular}{|l|ccc|}
\hline
Properties & Oak & Walnut & Birch \\ \hline
Material porosity $\varepsilon$ & $0.466$  & $0.634$ & $0.560$  \\
Dense-material density [kg/m$^3$]: $\rho^s/(1-\varepsilon)$  & $1486$ & $1661$   & $1777$\\
Darcy tensor [m$^2$]~\cite{debiagi2018predictive} & $10^{-14}$ & $10^{-14}$ & $10^{-14}$ \\
Thermal conductivity $\lambda^s$ [W/m/K]    & 0.13~\cite{flity2024thermal} & 0.04~\cite{forest1987wood}   & 0.18~\cite{forest1987wood}    \\ 
Biomass heat capacity $c^s$ [kJ/kg/K]~\cite{Gentile2017} & $1.5+0.001\,$T[K] & $1.5+0.001\,$T[K] & $1.5+0.001\,$T[K] \\
Char heat capacity $c^c$ [kJ/kg/K]~\cite{Gentile2017} & $0.7+0.0035\,$T[K] & $0.7+0.0035\,$T[K] & $0.7+0.0035\,$T[K] \\
    \hline
External heat transfer time $\tau_e = \rho^s c^s H / h$ & 60~s & 45~s & 55~s \\
Internal heat conduction time $\tau_i = \rho^s c^s H^2 / \lambda^s$ & 120~s & 300~s & 85~s \\ 
Characteristic Biot number $\textrm{Bi}=\tau_i/\tau_e$ & 2.0 & 6.6& 1.5 \\ \hline
\end{tabular}
\end{footnotesize}
\end{center}
\vspace{-0.2 in} 
\end{table*}

\subsection{Computational approach\label{sec:methods_sim}} \addvspace{10pt}
The experiments are complemented by numerical simulations using the BioSMOKE framework~\cite{Gentile2017}, which solves 3D volume-averaged conservation equations for mass, momentum, and energy in coupled solid and gas phases while dynamically adapting the mesh to resolve particle deformation. Particles are simulated as 3D cylindrical disks with diameters and height matching the experimental data, and a mean mesh cell size of $200$~$\mu$m validated through sensitivity analysis. The heat flux on lateral surfaces is set to retrieve lateral temperatures equal to the thermocouple measurements plus a constant offset tuned \emph{a posteriori} to $25$~K (see Fig.~S5). This offset corrects for probe fin cooling in calibration runs and its location on the upper surface away from the heated lateral surfaces. Top and bottom boundaries are kept adiabatic. Open boundaries allow for gas release, but the external gas phase is not modeled.

\begin{figure*}[h!t]
\centering
\vspace{-0.4 in}
\includegraphics[width=\textwidth]{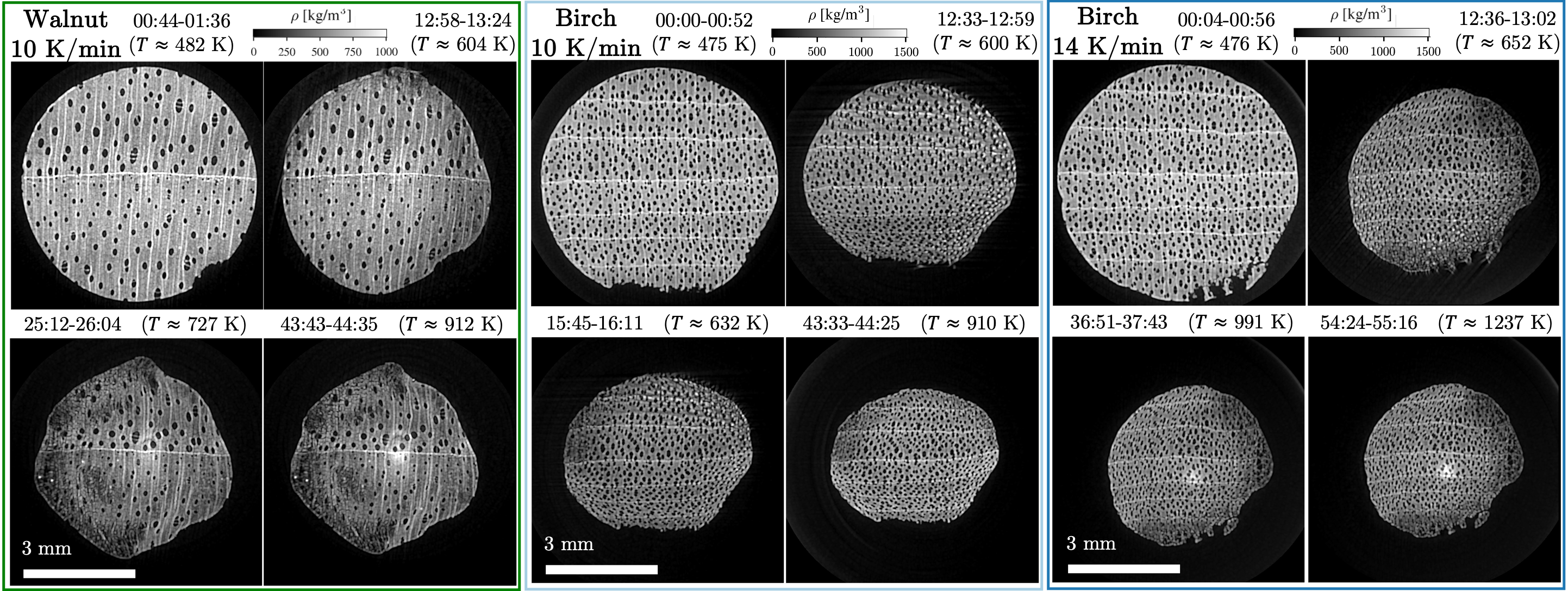}
\addvspace{2mm}\caption{\footnotesize Analysis of sample pyrolysis via \emph{in situ} $\mu$CT with $3.24$~$\mu$m spatial resolution, showing cross-sections of the measurements of density $\rho$ for walnut and birch biomass. The temperatures indicated are from thermocouple calibration.}
\label{FIG_SPOTTING}
\vspace{-0.2 in}
\end{figure*}

Reaction chemistry is computed with OpenSMOKE++~\cite{cuoci2015opensmoke++} using a recently developed mechanism derived from the CRECK kinetics of biomass pyrolysis. The global pyrolysis pathways summarized in Fig.~\ref{FIG_SETUP}\textbf{E} are modeled with 32 reactions with 27 solid-phase compounds releasing 29 gaseous species~\cite{debiagi2018predictive}. The solid residue is described as a mixture of pure carbon and lumped metaplastic compounds, representing oxygenated and hydrogenated groups attached to the carbonaceous matrix. The secondary pyrolysis is represented by the progressive tar and gas release from the metaplastic phase. Homogeneous gas reactions in the pores are not directly modeled. Based on material-specific analysis of the oak, birch, and walnut samples~\cite{boigne2021structural}, initial compositions are adjusted separately~\cite{debiagi2015extractives} to account for variations in moisture, cellulose, hemicellulose, three types of lignin, ash, tannin and triglyceride extractive (see supplementary Table~S2). Material properties are assumed isotropic, with values taken from measurements and literature (see Table~\ref{TAB_THERMO_PROP}).

\section{Results\label{SEC_RESULTS}} \addvspace{10pt}

\textit{In~situ} $\mu$CT measurements of density during biomass pyrolysis are shown in Figs.~\ref{FIG_OAK} and \ref{FIG_SPOTTING}. Intra-particle structural changes are captured in detail, resolving individual micro-pores and solid matrix deformation, though nano-scale porosity is not directly resolved. The biomass materials clearly differ at the pore-scale in density, pore size distributions, and grain structure. With oak, thermo-mechanical strain introduces cracks aligned with the wood grain, and fiber buckling contributes to pore deformation and crack propagation (red arrows in Fig.~\ref{FIG_OAK}). Comparing oak with walnut and birch (Fig.~\ref{FIG_SPOTTING}) reveals further material dependence: oak forms large cracks, walnut only minor ones, and birch none. Birch is a diffuse-porous hardwood whose uniform network of small pores vents volatile gases and distributes stresses, whereas the uneven pore structure of ring-porous oak and semi-ring porous walnut creates weak points and low-permeability regions that promote gas buildup and cracking. Supplemental movies S1--S4 provide a more complete overview of the $\mu$CT data.

\begin{figure}[ht!]
\centering
\includegraphics[width=185pt]{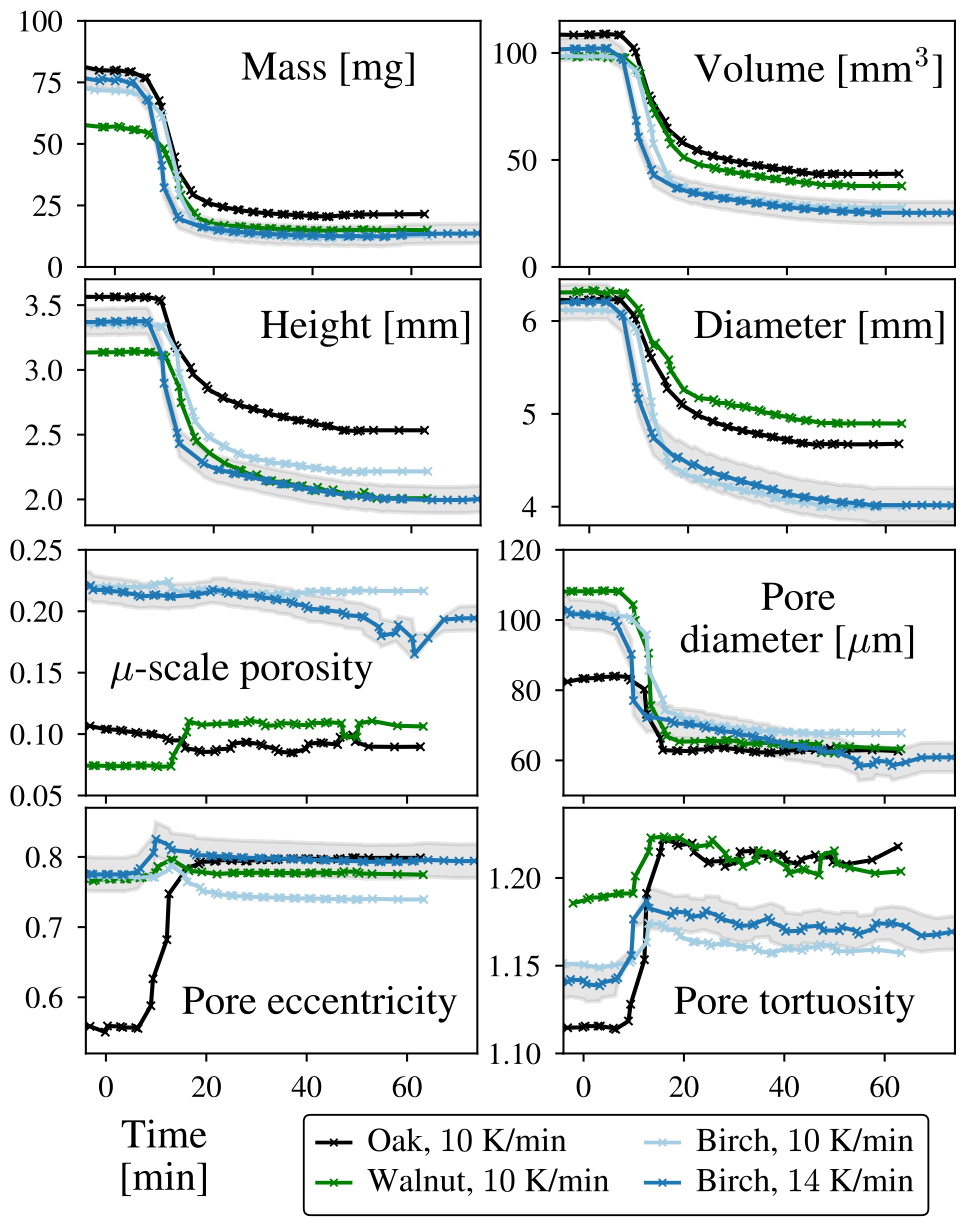}
\addvspace{2mm}\caption{\footnotesize Quantitative analysis during pyrolysis via $\mu$CT measurements, showing time-resolved characterization of the sample particle (mass, volume, height, and equivalent cylinder diameter), and the material pore-scale (resolved porosity, averages of pore diameter, eccentricity, and tortuosity). Experimental uncertainties are comparable across all runs; they are indicated with a shading for one dataset for readability.}
\label{FIG_RES_OVER_TIME1}
\vspace{-0.1 in}
\end{figure}

\begin{figure}[ht!]
\centering
\includegraphics[width=192pt]{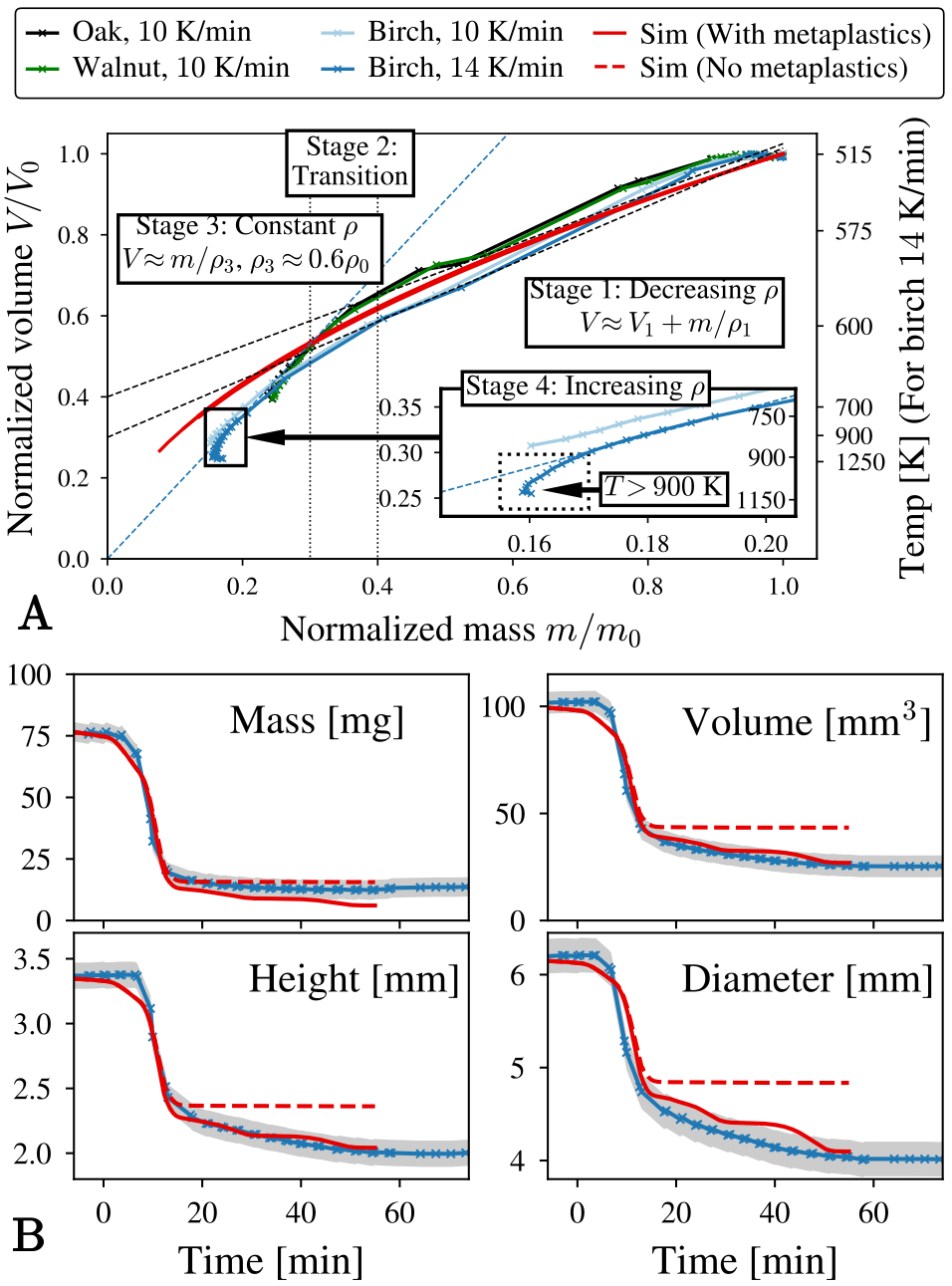}
\addvspace{2mm}\caption{\footnotesize Comparisons of experimental and simulations results. (\textbf{A}) Four stages of pyrolysis identified from the analysis of sample volume (TVA) and mass losses (TGA). The numerical results are nearly equal for the different materials and heating rates, such that the results collapse onto a single line. The black dashed lines are the best-fit lines in the first stage, and the blue dashed line indicates the best-fit line in the third stage for the birch heated at $14$~K/min. (\textbf{B}) Detailed comparisons for birch particle heated at $14$~K/min for simulations with and without metaplastics reactions.}
\label{FIG_RES_OVER_TIME2}
\vspace{-0.1 in}
\end{figure}

Despite a delay of $26$ to $150$~s between successive scans, pyrolysis dynamics is captured with enough photon accumulation for single-scan quantitative 3D density measurements without blurring or CT motion artifacts~\cite{boigne2022towards}. This demonstrates significantly improved contrast compared to previous ultra-fast $\mu$CT of biomass pyrolysis~\cite{Murai2020} at the same spatial resolution ($<10$~s scan time), highlighting the benefits of white-light mode operation with minimal filtering and optimized for quantitative analysis~\cite{Boigne2024}.

With this improved contrast, quantitative $\mu$CT provides time-resolved mass and volume data, enabling simultaneous TGA-TVA (Fig.~\ref{FIG_OAK}\textbf{A} and Fig.~\ref{FIG_RES_OVER_TIME1}). The measurements reveal anisotropic shrinkage unique to each wood type: walnut contracts nearly twice as much along the vessel direction as across it, birch exhibits the largest overall shrinkage, and oak the smallest, consistent with its mechanically stiff regions.

Beyond particle-averaged measurements, the $\mu$CT data is used to analyze individual pores and quantify micro-scale structural evolution during pyrolysis. Figure~\ref{FIG_RES_OVER_TIME1} reports time-resolved micro-porosity and mean pore diameter, eccentricity, and tortuosity. For oak, porosity remains nearly constant despite large structural deformations, as crack formation offsets pore expansion. Below $900$~K, pore properties change suddenly during primary devolatilization and evolve monotonically thereafter, except for tortuosity which spikes before gradually reducing. Above $900$~K, both porosity and tortuosity vary non-monotonically. Notably, birch porosity remains constant during primary devolatilization but decreases by nearly 20\% above $900$~K. These averaged properties integrate both pre-existing vessel pores and newly formed crack-induced voids, whose shifting relative population contributes to the non-monotonic trends. These non monotonic changes suggest the need for \emph{in situ} observation to fully capture the underlying phenomena.

\subsection{Stages of pyrolysis\label{ssec:pyro_stages}} 
To further compare the experimental and simulation results, we first identify four stages of mass and volume losses for our conditions (Fig.~\ref{FIG_RES_OVER_TIME2}\textbf{A}): (i) rapid primary devolatilization with decreasing density between $545$–$610$~K, (ii) transition over $610$–$650$~K to secondary pyrolysis with more volume losses, (iii) continued devolatilization at nearly constant density from $650$–$900$~K, and (iv) high-temperature char devolatilization above $900$~K where density rises.

The first stage is cellulose breakdown, completed by $650$~K at our heating rates~\cite{Anca-Couce2012}. During the transition, the preferential formation of gaseous volatiles over solid char at higher temperatures increases the ratio of volume to mass loss~\cite{DiBlasi2001, Anca-Couce2016}. Mass loss plateaus above $650$~K, where char decomposition releases H$_2$ and CO. Beyond $900$~K, density increases despite continued mass loss, implying more condensed reaction products. At $82\pm5$\% mass loss, density is $12\pm3$\% higher at $1240$~K than at $900$~K, revealing a distinct pyrolysis stage during which material density increases. At these temperatures, previous TGA experiments combined with mass spectrometry or FTIR~\cite{Widyawati2011, Collard2014} have proposed new formation pathways for dense polycyclic aromatic hydrocarbons via dehydrogenation and substitution reactions releasing CH$_4$ and CO. The density increase measured here could represent supporting evidence, although char graphitization and collapse of unresolved nano-scale porosity could also contribute.

\subsection{Comparisons with simulations\label{ssec:comp}} 
The experimental data is compared to 3D particle-resolved simulations. Figure~\ref{FIG_RES_OVER_TIME2}A shows that overall trends and shrinkage are well captured, but residual masses are under-predicted by more than a factor of two for all cases. The model also fails to capture the material-specific variability in mass-volume losses, especially in the later pyrolysis stages. In Fig.~\ref{FIG_RES_OVER_TIME2}B, we compare the $\mu$CT data for birch heated at $14$~K/min with simulations that consider and neglect the metaplastic reactions modeling secondary pyrolysis. Neglecting them reduces the residual mass error, but they are necessary to predict volume losses in secondary pyrolysis. Similar trends are observed at $10$~K/min (see supplementary Figs.~S2--S4).

These results highlight modeling deficiencies in simultaneously predicting mass and volume, with underlying accuracy trade-offs. Most kinetic models are tuned to TGA data only~\cite{debiagi2018predictive, debiagi2024cellulose}, making them prone to overfitting mass data, whereas quantitative $\mu$CT provides joint TGA-TVA as more robust benchmark data. A likely contributor to the residual mass mismatch is the present model's indirect treatment of secondary pyrolysis through solid-phase metaplastic kinetics, which precludes direct resolution of secondary tar cracking in the pores and limits sensitivity to anisotropic volatile transport effects. Further comparisons with other pyrolysis models such as flash distillation chain statistics or chemical percolation devolatilization (CPD) could help further identify the origins of these deficiencies~\cite{pielsticker2024comparative}. The model also neglects detailed mineral matter, particularly alkaline metals with known catalytic effects on volatile cracking~\cite{BrennanPecha2019, Anca-Couce2016}. Recent improvements to the CRECK cellulose submodel~\cite{debiagi2024cellulose, caraccio2026multiphase} may have further implications for secondary pyrolysis predictions. Uncertainties from sample positioning and temperature calibration could also contribute to the reported mismatch.

\subsection{Regimes of internal physico-chemical coupling\label{ssec:res_regimes}} 

For particles such as firebrands~\cite{manzello2020role} that are large enough to feature non-uniform reaction rates and temperature distributions, the pyrolysis is governed by the coupling between chemistry and intra-particle transport. The Damköhler number $\text{Da} = \tau_t/\tau_c$ characterizes this coupling, where $\tau_c$ is the chemical time-scale and $\tau_t \approx 160$~s is the transport-limiting timescale (conduction here, see Table~\ref{TAB_THERMO_PROP}). For $\text{Da} < 0.1$, it is often assumed that reactions occur uniformly~\cite{Pyle1983, Corbetta2014, Pecha2021}, and intra-particle transport can be neglected.

\begin{figure}
\centering
\includegraphics[width=192pt]{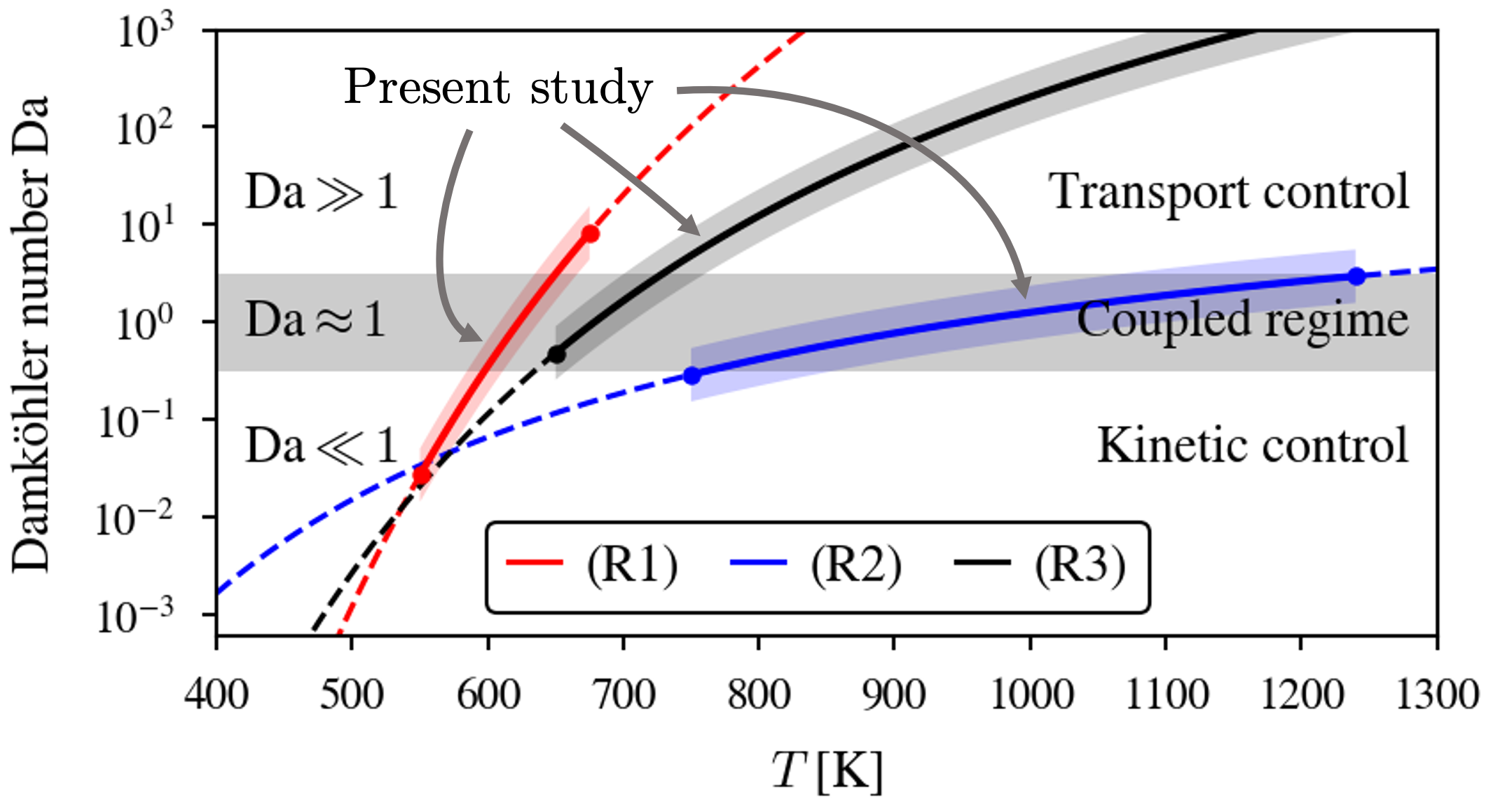}
\addvspace{2mm}\caption{\footnotesize Theoretical predictions of the regimes of intra-particle coupling for the global reactions considered. The full lines indicate the estimated ranges of temperature over which each reaction evolves in the present study (grey arrows), indicating the regimes for each reaction. The regimes are characterized by the Damköhler number 
$\textrm{Da}=\tau_t/\tau_c$ where $\tau_t$~s is at internal conduction time evaluated from the calibration data with shadings indicating the range from $85$~s (birch) to $300$~s (walnut, see Table~\ref{TAB_THERMO_PROP}), and $\tau_c=1/k$ is the chemical time-scale with temperature dependence controlled by the Arrhenius reaction rate $k_i = A_i \exp(-E_{a,i}/RT)$. The kinetic values used from the literature are: $A_1 = 4.38 \times 10^9$~s$^{-1}$, $E_{a,1}= 141.2$~kJ/mol~\cite{DiBlasi2001}, $A_2 = 6.55 \times 10^{-1}$~s$^{-1}$, $E_{a,2}= 36.7$~kJ/mol~\cite{Cordero1990}, and $A_3 = 9.55 \times 10^4$~s$^{-1}$, $E_{a,3}= 93.3$~kJ/mol~\cite{Boroson1989}.}
\label{FIG_KINETICS_PLOT}
\vspace{-0.1 in}
\end{figure}

For the present conduction times, Fig.~\ref{FIG_KINETICS_PLOT} shows the temperature-dependence of the Damköhler numbers for reactions (R1), (R2), and (R3), computed using literature chemical kinetics~\cite{DiBlasi2001, Cordero1990, Boroson1989}. The coupling regime for each reaction is determined by the temperature range over which it occurs, estimated \emph{a posteriori} from the $\mu$CT results, though this determination carries large uncertainties from evaluating the intra-particle temperature field. This time-scale analysis predicts that heating rates are high enough and samples sufficiently large for reactions to span regime transitions, such that neither the kinetic nor transport asymptotic limits alone are sufficient to model pyrolysis accurately~\cite{ringel2025carbon, jiang2025heating}.

\begin{figure*}[h!t]
\centering
\vspace{-0.4 in}
\includegraphics[width=\textwidth]{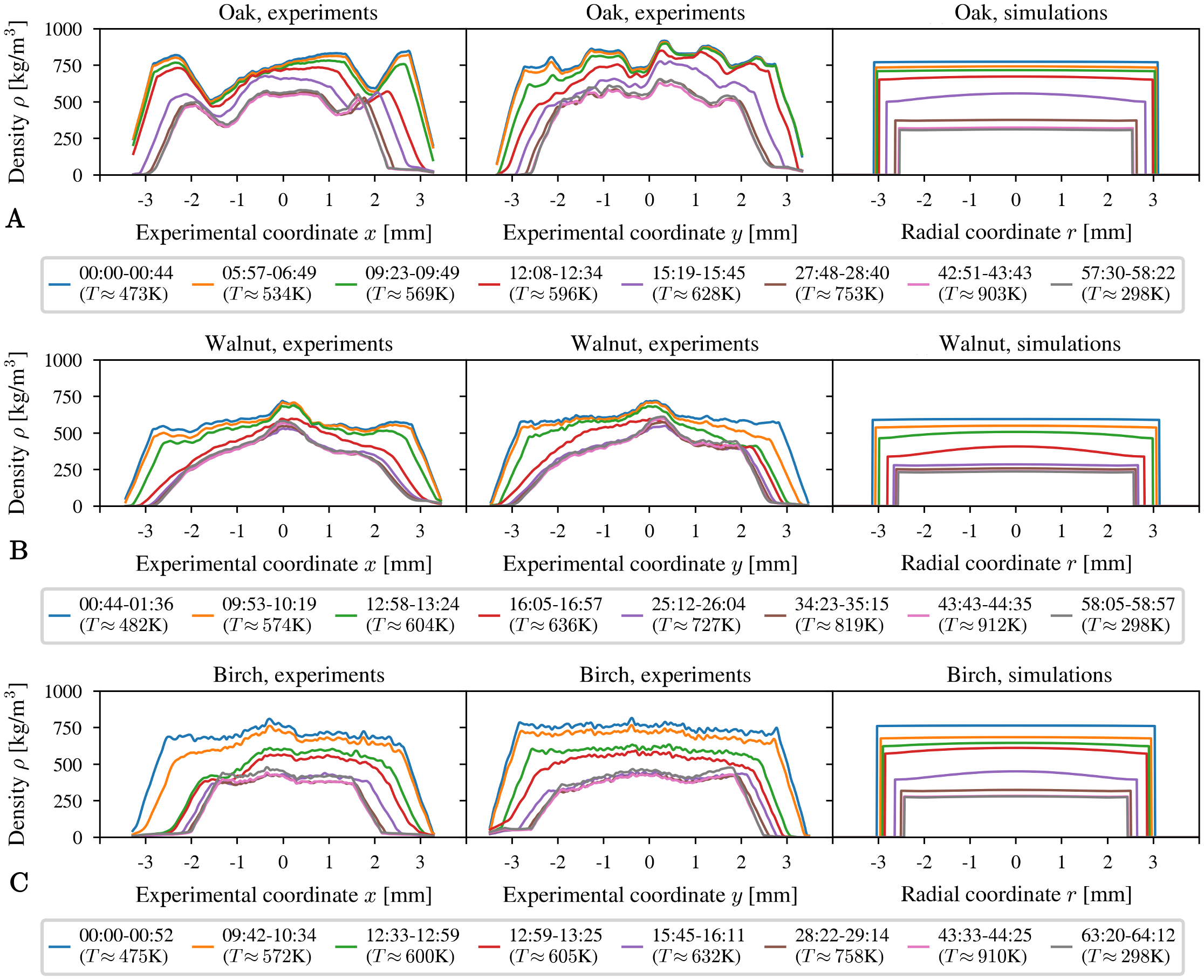}
\addvspace{2mm}
\caption{\footnotesize Spatial profiles of density from the experiments and simulations. Experimental results (left and central panels) are provided for heating rates of $10$~K/min with (\textbf{A}) oak, (\textbf{B}) walnut, and (\textbf{C}) birch biomass. The 3D experimental results are averaged with a kernel of $0.5$~mm width before extracting the 1D profiles across the grain along the lines $y=0$ (along the axis $x$ of wood growth) and $x=0$ (perpendicular to the axis of wood growth) in the time-dependent plane of constant $z$ taken at a quarter of the sample height from the bottom. The numerical results (right panels) are from the simulations without metaplastics.}
\label{FIG_DENSITY_LINES}
\vspace{-0.2 in}
\end{figure*}

In agreement with these results, the $\mu$CT data and transverse profiles in Fig.~\ref{FIG_DENSITY_LINES} show that oak and walnut pyrolyze non-uniformly from the outer surfaces inward, while birch at $10$~K/min pyrolyzes nearly uniformly, consistent with its faster conduction time and kinetically-controlled regime. The 3D simulations predict internal temperature differences of $10$~K (birch) to $30$~K (walnut) during pyrolysis, and non-uniform density profiles visible in Fig.~\ref{FIG_DENSITY_LINES} at temperatures of $630\pm20$~K, independently of material. Specifically, the simulations predict a similar extent of pyrolysis homogeneity across all materials despite accounting for material-specific density and thermal conductivity, and therefore fail to capture the regime differences between more homogeneous birch pyrolysis, and less homogeneous oak and walnut pyrolysis. Again, this insensitivity is likely a direct consequence of the model indirect treatment of secondary pyrolysis. Without explicitly resolving gas-phase homogeneous reactions within the pores, the present model cannot capture material-specific differences related to anisotropic volatile transport.

These findings suggest that current models can overlook sensitive regime transitions, in part because intra-particle regimes remain hard to characterize \emph{in situ}. In simple time-scale analysis, uncertainties compound from determining temperature \emph{a posteriori} and lumping multi-step kinetics. Furthermore, characteristic heat transfer times are often computed neglecting the heat of reaction~\cite{BrennanPecha2019, Pecha2021, Anca-Couce2016}. In fact, the endothermic heat of primary pyrolysis ($\Delta H_1 \approx 300$~kJ/kg~\cite{Rath2003}) can increase the effective heat transfer time fivefold, when considering not only the sensible enthalpy $c^s\Delta T/\Delta t$ but also the chemical enthalpy $\Delta H_1 \dot{m}/m$. 

Accounting for this factor in Fig.~\ref{FIG_KINETICS_PLOT} would shift all curves upwards by an amount slightly exceeding the shown $3.5\times$ spreading from birch to walnut (colored shadings). It leaves the coupling regime classification for primary pyrolysis unchanged but potentially reclassify secondary pyrolysis as transport-controlled, a regime for which the $\mu$CT data provide no direct observational evidence. More generally, accounting for the pyrolysis heat could also explain why anisotropy can double consumption times at low $\text{Da}\sim 0.2$~\cite{Pecha2021}. At minimum, distinguishing timescales of primary devolatilization from secondary pyrolysis appears critical for accurate regime identification. Last, comparing density profiles along and across the direction of the wood growth (Fig.~\ref{FIG_DENSITY_LINES}) clearly show that significant material anisotropy is lost by the present isotropic model.

\section{Conclusions\label{SEC_DISCUSSION}} \addvspace{10pt}

This work presents joint experimental and computational results from combined $\mu$CT and particle-resolved simulations to enable a quantitative \emph{in situ} analysis of intra-particular pyrolysis under well controlled conditions. The $\mu$CT contrast and accuracy are largely improved in comparison to existing literature work~\cite{Murai2020}. The results show that, despite fast acquisition rates, synchrotron $\mu$CT can retrieve quantitative measurements of porous micro-scale, even for low-radiodensity organic materials. The proposed approach can be used to characterize pore-scale structural changes in detail, enable simultaneous TGA-TVA measurements, and conduct more systemic comparisons with detailed simulations~\cite{caraccio2026multiphase, Marti-Rossello2021, Lachaud2017}.

Above $900$~K, we measure an increase in mean particle density, which could support previously proposed chemical pathways for dense polycyclic aromatic hydrocarbons~\cite{Widyawati2011, Collard2014}. Experiments show clear differences across three different biomass materials, whereas detailed simulations provide nearly identical predictions, despite accounting for differences in chemical compositions and thermophysical properties. The datasets presented with joint TGA-TVA represent quantitative benchmarks for model evaluations, and are therefore particularly relevant to improve models of particle pyrolysis regimes~\cite{BrennanPecha2019, Pecha2021}, the underlying chemical processes~\cite{debiagi2018predictive, debiagi2024cellulose}, and the anisotropic shrinkage and deformation~\cite{caraccio2026multiphase, Lahouze2025, li2023comprehensive}. The detailed experimental and simulation data is shared online through the Stanford Digital Repository at \url{https://doi.org/10.25740/tf116vx1269}. 

Although this work focuses on biomass conversion prior to oxidation, it represents a first step towards general experiments of chemically reacting particles, such as smoldering or char combustion, or industrial calcination and sintering. Furthermore, X-ray diagnostics is increasingly being extended to simultaneous gas-phase measurements~\cite{Sakurai2016, hansen2017, Boigne2024}. These developments, together with combined species-selective exhaust and gas component analyses~\cite{debiagi2024cellulose, topperwien2025burn}, the capabilities of upcoming fourth generation synchrotron sources, and optomechanical breakthroughs for CT at sub-$10$-nm resolution~\cite{de2021fast}, are enabling multiphase investigation of solid and volatile phases at previously inaccessible spatial and temporal resolution~\cite{ringel2025carbon, jiang2025heating}.

\acknowledgement{CRediT authorship contribution statement} \addvspace{10pt}

{\bf EB}: designed research, performed experiments, analyzed data, wrote paper. {\bf MMA}: performed simulations, analyzed data, edited paper. {\bf CF}: performed experiments. {\bf ERT}: performed experiments. {\bf DYP}: designed research, performed experiments, reviewed paper, supervision. {\bf AAM}: performed experiments, reviewed paper, supervision. {\bf HSB}: performed experiments. {\bf FP}: reviewed paper, supervision. {\bf CS}: performed simulations, reviewed paper, supervision, funding. {\bf MI}: designed research, performed experiments, edited paper, supervision, funding.

\acknowledgement{Declaration of competing interest} \addvspace{10pt}
The authors declare that they have no known competing financial interests or personal relationships that could have appeared to influence this work.

\acknowledgement{Acknowledgments} \addvspace{10pt}
This research was supported by the Moore Foundation, the US NSF (Award No. CBET-1800906), and the US DOE, Office of Basic Energy Sciences (Award No. DE-SC0026165). It used resources of the Advanced Light Source, which is supported by the Director, Office of Science, Office of Basic Energy Sciences, of the US DOE (Contract No. DE-AC02-05CH11231), and operated for the DOE Office of Science by Lawrence Berkeley National Laboratory. The work done by CS and MMA was performed under the auspices of the US DOE by Lawrence Livermore National Laboratory under contract no. DE-AC52-07NA27344 and was supported by the LLNL-LDRD Program under project no. 23-LW-007.

\ifdefined\final
\footnotesize
\baselineskip 9pt
\fi

\clearpage

\pagestyle{empty}

\bibliographystyle{proci}
\bibliography{science_template}

\end{document}